\documentstyle[11pt,paspconf,psfig]{article}

\begin{document}

\title{A Self-consistent Method for Estimating Photometric Redshifts}
\author{Bahram Mobasher$^1$ \& Paola Mazzei$^2$}
\affil{$^1$ Astrophysics Group, Blackett Laboratory, Imperial College,
    Prince Consort Road, London SW7 2BZ, U.K\\
$^2$ Osservatorio Astronomico, Vicolo dell'Osservatorio, 5, 
35122 Padova, Italy}

\begin{abstract}
A new method is developed for estimating photometric redshifts, using
realistic template SEDs, extending over four decades in wavelength
(i.e from 0.05 $\mu m$ to 1 mm). The SEDs are constructed for four
different spectral types of galaxies (elliptical, spiral, irregular
and starburst) and satisfy the following characteristics:
a). they are optimised to produce the observed colours
of galaxies at $z\sim 0$; b). incorporate the chemo-photometric
spectral evolution  
of galaxies of different types,    
in agreement with observations; c). allow treatment of dust contribution
and its evolution with redshift, consistent with the spectral evolution model; 
d). include absorption and re-emission of radiation by dust and hence, 
realistic estimates of the far-infrared radiation;
e). include correction for inter-galactic absorption by Lyman continuum
and Lyman forest. 

Using these template SEDs, the photometric redshifts are estimated to an 
accuracy of $\Delta z = 0.11$. The degeneracy in the method and hence, 
the redshifts and spectral types,   
are explored via Monte Carlo simulations.  
The consistency of the technique is demonstrated by estimating
photometric redshifts to a UV selected sample of HDF galaxies
and investigating if the     
statistical results (ie. luminosity densities) are
in agreement with other independent studies.

\end{abstract}


\section{Introduction}

The subject of photometric redshift measurement (i.e. estimating
redshifts to galaxies from multi-waveband photometric observations) 
has recently been revitalised, due to a rapid increase in the number of 
existing multi-waveband galaxy surveys.  
This provides a fast way for estimating redshifts to a large number 
of galaxies, with some trade-off in their accuracy. Moreover, in the case
of faint galaxies ($I > 25$ mag.), where spectroscopic redshifts
are more difficult to measure even with the largest telescopes,  
photometric redshift technique is the {\it only} way to secure their 
redshifts. 

For a given galaxy, the photometric redshift can be estimated by
comparing its observed SED with a set of template SEDs, corresponding to
different galaxy types and shifted to different redshifts, 
allowing for the galaxy evolution with look-back time. The redshift and
spectral type associated with the template SED which is the closest 
to the observed SED will then be assigned to that galaxy. Therefore,
the crucial problem is the choice of the template SEDs and their
behaviour at high redshifts. 
There are two general ways for adopting the template SEDs, as discussed
below; 

\noindent {\bf a).} Empirical templates: in this case one uses the observed SEDs for 
different types of galaxies. 
The problem here is that there are not enough 
information about the SEDs for different classes of galaxies
at different redshifts (particularly at high redshifts). 
Therefore, incorporating the spectral evolution
of galaxies to the model SEDs of different types is difficult and uncertain.

\noindent {\bf b).} Synthetic templates: uses model SEDs for different 
spectral types of
galaxies, shifted in redshift space, assuming evolutionary population 
synthesis (EPS) models. The main problem here is to construct
realistic model SEDs at different redshifts by constraining the 
evolutionary models.  

In the present study, we adopt a combined approach to self-consistently 
model the
template SEDs at different redshifts, in agreement with observations. 
In particular, the effect of dust and its evolution with redshift
is included in the SEDs, consistent with the EPS models,
and optimised using observations
at different look-back times.

\section{The Photometric Redshift Technique- A New Approach}

A large set of chemo-photometric Evolutionary Population 
Synthesis (EPS) models with different input parameters
are developed, each accounting for both the local properties and
evolutionary behavior of different types of galaxies$^{1,2}$. 
For a given spectral type of galaxies, 
the model parameters are normalised to those
at $z=0$, by fitting them to the local observed SED of their respective
type. The evolutionary behaviour (i.e. input parameters) 
of the EPS models (which also include contribution from dust) are then 
constrained
by estimating the photometric redshifts ($z_{phot}$) 
to a {\it calibrating} sample
of 73 galaxies with available spectroscopic redshifts ($z_{spec}$) 
and minimising the
{\it rms} scatter between them. The calibrating sample is adopted   
from the Hubble Deep Field (HDF) so that the galaxies will have detection
in, at least, four passbands, including UV which is crucial for any
photometric redshift determination. 
The procedure is summarised in the flow chart in Figure 1. 

The parameters in the EPS models, constrained by observations, consist of
the IMF, formation redshift ($z_{form}$), local star formation rate 
($\psi_0$) and the time evolution of the star-formation rate, as parametrised
by $n$ (see Figure 1). Once these parameters are constrained, 
the evolution of the optical depth ($\tau$) is determined
directly from the gas metallicity, $Z_{gas}$, and the fractional
mass of gas which takes part in star formation ($f_g$), assuming that
the gas-to-dust ratio is proportional to the gas metallicity$^{1,2}$. 
For galaxies with $z > 2$ in the calibrating sample, the effect of 
Lyman continuum and Lyman forest absorption is estimated$^3$  
The fluxes for these objects are subsequently corrected
for inter-galactic absorption before they are used to constrain the EPS models.
The template SEDs cover a range in wavelength from 0.05 $\mu m$ to 1 mm. 

The sensitivity of the result to the EPS model parameters 
are shown in Table 1.   
The optimum {\it rms} scatter 
obtained on the $z_{phot}$ vs. $z_{spec}$
plane (Fig. 2) is 0.11 (model 4 in Table 1), 
which is found using template SEDs for 
four different types of galaxies (elliptical, spiral, 
starburst and irregular) and only four passbands (UVRI). 
This relation for the calibrating sample
is presented in Figure 2, with its {\it rms} scatter taken 
as the error in our photometric redshifts. 

\begin{figure}
\centerline{\psfig{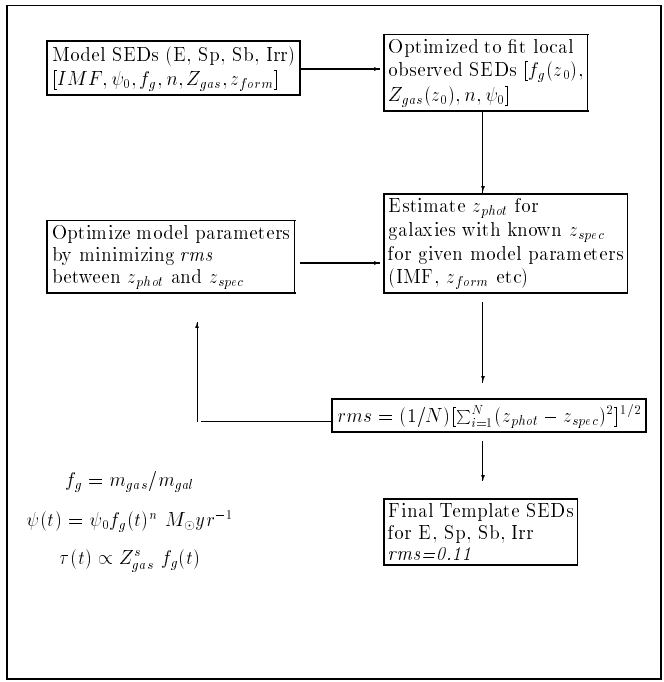}}
\caption{
Flow chart showing the new procedure for finding 
template SEDs }
\label{diagnostic}
\end{figure}

\begin{figure} 
\centerline{\psfig{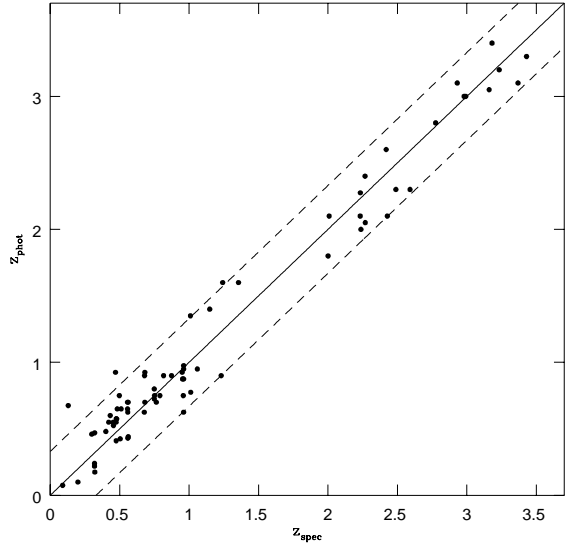}}
\caption{$z_{phot}\ vs.\ z_{spec}$ for the calibration sample of 73 HDF galaxies.
Dashed lines correspond to the 3$\sigma$ limits.}
\label{diagnostic}
\end{figure}

\begin{table*}
\pagestyle{empty}
\caption{Sensitivity of the {\it rms} scatter between the photometric
and spectroscopic redshifts on the EPS model parameters (i.e. template SEDs)}   
\begin{tabular}{ccccccccc}
  Model & IMF & $m_l$ & Number of & \multicolumn{4}{c}{$z_{form}$} & 
{\it rms} \\
  &        &    &       Templates & E & Sp & Sb & Irr  &\\
  &        &    &       &           &   &    &    &          \\ 
 1 & Salpeter  &      0.01 &      4  &  5&  5&    5 &   1 &     0.18 \\
 2 & "  &      0.01 &      3  &  5&  5&    5 &   -- &    0.26 \\ 
 3 & "  &      0.01 &      3  &  5&  2&    5 &   -- &    0.24 \\ 
 4 & "  &      0.01 &      4  &  5&  2&    5 &   1  &    0.11 \\ 
          &           &         &   &   &    &    &         \\ 
 5 & Salpeter  &      0.10 &      4  &  5&  5 &    5 &   1 &     0.27 \\ 
 6 & "       &      0.10 &      3  &  5&  5 &   5 &   -- &    0.38 \\ 
 7 & "       &      0.10 &      3  &  5&  2 &   5 &   -- &    0.37 \\ 
 8 & "       &      0.10 &      4  &  5&  2 &   5 &   1  &    0.22 \\ 
          &           &         &   &   &    &    &         \\  
 9 & Scalo     &      0.10 &      4  &  5&  5 &   5 &   1  &    0.28 \\ 
 10 & "     &      0.10 &      3  &  5&  5 &   5 &   -- &    0.52 \\ 
 11 & "     &      0.10 &      3  &  5&  2 &   5 &   -- &    0.45 \\ 
 12 & "     &      0.10 &      4  &  5&  2 &   5 &   1  &    0.25 \\ 
\end{tabular}
\end{table*} 

\section {Degeneracy in Photometric Redshifts }

A source of uncertainty in estimating the photometric redshifts and
spectral types of galaxies is the possibility that the models might be
degenerate (i.e. different template SEDs producing the same result). 
Furthermore, the photometric errors in the observed SEDs
are likely to affect the final estimate of both the photometric redshift
and the spectral type of their respective galaxy. 

To investigate the uniqueness of the photometric redshift 
solutions in this study, a 
Monte Carlo simulation is performed. A simulated catalogue is generated
to resemble the observed, UV selected HDF survey, with UBVI magnitudes, 
known redshifts ($z_{input}$) and spectral types (a UV selected catalogue
is used here to ensure the availability of UV data which are crucial for
any photometric redshift measurement). The galaxies in this catalogue are
randomly selected to have SEDs similar to the synthetic SEDs for any
of the four types
of galaxies (E, Sp, Sb, Irr) considered here, shifted in redshift space. 
Random Gaussian noise, resembling photometric errors are then added to the
simulated SEDs. The simulated catalogue contains 273 galaxies (the same as
the complete UV-selected HDF catalogue), has a magnitude limit of 
U=27 mag., an apparent UV magnitude distribution similar to the observed
catalogue and a redshift limit of $z=2.5$ (maximum redshift allowed in 
a UV selected survey).

The photometric redshift code was then used to predict the redshifts ($z_{output}$)
and spectral types of individual galaxies in the simulated catalogue
and to compare them with their {\it input} values. 
The $log(z_{output}/z_{input})$
distribution (Fig. 3) shows a distinct peak at zero, indicating that
the redshifts for the simulated galaxies are well re-produced within
$\Delta z \sim 0.11$. 

As far as the predicted spectral types are
concerned, we recover the type classification for ellipticals 
in the input catalogue by $100\%$
(i.e. no mis-identification of ellipticals). Regarding  
the spirals, irregulars and starbursts, we can recover, respectively, 
$79\%$, $85\%$ and $71\%$ of the spectral types of galaxies
in the input catalogue.

\begin{figure}
\vspace*{-4cm}
\centerline{\psfig{figure=fig3_4.enps,width=0.8\textwidth,angle=0}}
\caption{log ($z_{output}/z_{input}$) results from simulation. 
The peak at zero indicates that the redshifts in the
simulated catalogue are re-produced}
\caption{ Changes in the UV luminosity densities with redshift,  
using photometric redshifts for the UV selected HDF survey (filled circles).
Compared with estimates from
Lilly et al (1996)-(crosses); Connolly et al (1998)-(triangles); 
Madau et al (1996)-(filled squares). 
}
\label{diagnostic}
\end{figure}

\section {Consistency Test}

In order to explore the consistency of the above procedure, we apply it
on a complete sample of UV selected galaxies from the HDF, to estimate
their photometric redshifts. Using this redshift catalogue, we then 
predict the physical quantities (i.e. the 
luminosity density) in redshift intervals
and compare them with similar measurements from
other independent studies. A UV selected survey is particularly useful 
for such comparison because the UV light is heavily affected
by dust in galaxies$^4$, allowing a test
of the estimated extinction corrections. Moreover, such surveys do not
contain objects with $z > 2.5$ and hence, are less affected (then deep
surveys in other bands) by uncertainties in correction for 
Lyman continuum absorption.

The UV luminosity densities are estimated  in redshift intervals, 
using the photometric redshift catalogue. 
The results are compared with similar measurements from other, 
independent, studies (Fig. 4) and 
show excellent agreement. This indicates that the statistical results, 
based on the present photometric redshift technique, are consistent
with other studies. 

\section {Summary and Conclusion}

A new method is presented to estimate photometric redshifts of galaxies. 
The main improvement in this study is to develope more realistic 
template SEDs over a large range in redshift and
in agreement with the observational data. The SEDs cover a range
in wavelength from 0.05 $\mu m$ to 1 mm, with  
their main characteristics as follows:
a). they are optimised to produce the observed colours
of galaxies at $z\sim 0$; b). incorporate the 
chemo-photometric evolution 
of galaxies of different types (i.e. stellar luminosities
for different metallicities), 
in agreement with observations; c). allow a self-consistent 
treatment of dust contribution
and its evolution with redshift; 
d). include absorption and re-emission of radiation by dust and hence, 
realistic estimates of the far-infrared radiation;
e). include correction for inter-galactic absorption by Lyman continuum
and Lyman forest. 

Using a calibration sample of 73 galaxies with available spectroscopic 
redshifts, the evolutionary model
parameters are optimised to get the minimum scatter between the 
photometric and spectroscopic redshifts in the calibrating sample.
Using four passbands (UVRI), the photometric redshifts are estimated to
an accuracy of $\Delta z =0.11$. 

The method is applied to a UV selected sample of HDF galaxies. The
UV luminosity densities, estimated in redshift
intervals, are in excellent agreement with similar measurements 
from other, independent, 
surveys.

\end{document}